\documentclass[useAMS,usenatbib]{mn2e}
\usepackage{epsfig}
\usepackage{lineno}
\usepackage{lscape}
\usepackage{graphicx}
\usepackage{amssymb}
\usepackage{rotating}
\usepackage{natbib}

\title[New variabilities from IGR J17091--3624]{Evidence of two unique variability classes from IGR J17091--3624}
\author[Pahari et al.]{Mayukh Pahari$^{1}$\thanks{E-mail: mp@tifr.res.in}, Sudip Bhattacharyya$^{1}$, J S Yadav$^{1}$ and S K Pandey$^{2}$ \\
$^1$ Tata Institute of Fundamental Research, Homi Bhabha Road, Mumbai 400005, India\\
$^2$ School of Studies in Physics \& Astrophysics, Pt. Ravishankar Shukla University, Raipur, Chhattisgarh 492010, India}
\begin{document}

\date{}

\pagerange{\pageref{firstpage}--\pageref{lastpage}} \pubyear{2012}

\maketitle

\label{firstpage}

\begin{abstract}

IGR J17091--3624 is the second black hole X-ray binary after GRS 1915+105, 
which showed large and distinct variabilities.
The study of these variability classes can be useful to understand the accretion-ejection mechanisms
of accreting black holes, and hence to probe the strong gravity regime. We report the
discovery of two new variability classes (C1 and C2) from IGR J17091--3624 from the 2011
outburst {\it Rossi X-ray Timing Explorer} data. These unique 
classes will be useful to have complete details about the source, and to learn new aspects
about variabilities. For examples, the C1 class shows that the intensity and period of
oscillations, energy spectrum and power spectrum can clearly evolve in tens of seconds. Moreover,
in such a small time scale, soft-lag becomes hard-lag. the C2 class shows that the variability
and the nonvariability can occur at similar energy spectrum, and a soft state is not required
for variability to happen.
 \end{abstract}  


\section{Introduction}\label{Introduction}

In its 16 years of service, the {\it Rossi X-ray Timing Explorer} ({\it RXTE}) detected around 
$\sim$40 black hole X-ray binaries (BHXB; \citep{Rodriguezetal2011,remandmcc2006}), 
among which GRS 1915+105 is the most extraordinary and prolific in terms of its unique
variabilities \citep{mirandrod1994,Eikenberryetal1998,Yadavetal1999,Bellonietal2000}. 
This source showed about 12 types of distinct classes, whereas no such distinct fast variabilities 
in the intensity were observed from any other BHXBs \citep{Bellonietal2000,Yadavetal1999}. 
Therefore, while a few limited X-ray states (e.g., low hard state, high soft state etc.) 
and transitions among them are studied for other BHXBs in order to understand the accretion-ejection
mechanisms \citep{remandmcc2006,Homanetal2001,Bellonietal2000}, GRS 1915+105 provides a unique laboratory to probe the
inflow-outflow mechanism in many ways \citep{Yadav2006}. However, in order to use X-ray variabilities from
GRS 1915+105 as a tool, one needs to understand them adequately. Although, it is possible
that these variabilities originate from accretion disc instabilities, definite conclusions
could not be made due to the lack of another source showing similar properties \citep{Bellonietal2000}. 

Recent outburst of another black hole X-ray binary IGR J17091--3624 (e.g., \citep{Krimmetal2011,
Rodriguezetal2011}) has shown several X-ray variability classes \citep{Altamiranoetal2011b} similar to some of the classes (e.g., $\nu$, $\rho$, $\alpha$,
$\beta/\lambda$, $\mu$) observed from the GRS 1915+105 \citep{Bellonietal2000}. Observations
of a given class with different intensities, energy spectra and power spectra from two
different sources can be very useful to understand the physics of this class by constraining its models.
For example, \citet{Paharietal2011} studied the $\rho$-class properties of IGR J17091--3624 in detail,
and compared them with those of GRS 1915+105. In this Letter, we report the discovery of two
types of variabilities, which have never been observed from any black hole X-ray binary.
Using spectral and timing studies, and from the evolution and repeatability of these
variabilities, we establish their uniqueness, and define them as new classes (C1 and C2).
The unique properties of C1 and C2 can be useful to understand the variabilities from
IGR J17091--3624 and GRS 1915+105 in general, and hence to probe the accretion-ejection mechanisms.

\section{Data analysis}\label{Dataanalysis}

We analyze all the Good Xenon mode {\it RXTE} proportional counter array (PCA) data
(203 obsIds; up to November 18, 2011) from the 2011 outburst of the transient 
black hole X-ray binary IGR J17091--3624. We examine light curves from all these
obsIds, and find two new classes of variabilities (see \S~\ref{Results}), which
were never observed from any black hole X-ray binary, not even from GRS 1915+105.
We call them the C1 class and the C2 class which are discussed in details in next section.

In order to check whether these are really new classes, as well as to understand their 
properties, we compute hardness-intensity diagram (HID; intensity versus hard colour), colour-colour diagram (CD; soft colour
versus hard colour), power spectrum and time delay between two energy ranges.
The hard colour is defined as the ratio of the background-subtracted count rate in $12-60$ keV
to that in $2-5$ keV. The soft colour is the ratio of the background-subtracted 
count rate in $5-12$ keV to that in $2-5$ keV. The intensity is the background-subtracted
count rate in $2-60$ keV. The colour value depend on the energy spectrum, and
provide a convenient way to track the spectral evolution.  
Since proportional counter unit 1 (PCU1) and unit 2 (PCU2) were operating during observations 
of C1 and C2 classes, we use both of them. 
We compute a power density spectrum (PDS) from a 1024 s interval with a Nyquist
frequency of 100 Hz in the entire PCA energy range, using the standard
Fourier transform and considering the normalization of \citet{Miyamotoetal1991}.
The PDS is geometrically re-binned by a factor of 1.05, in order to suitably reduce 
the error throughout the frequency range. In order to find a plausible time delay between 
the variabilities in different energy bands, we calculate the cross correlation coefficient 
between light curves in two energy bands (see, for example, \citet{Srirametal2010}). The shift of the peak of the cross correlation function from the origin provides 
the delay time. Since we are unaware of the time delay at the beginning, we started with 5 ms bintime. This choice ensures that the binsize is sufficiently smaller than the delay time. We cross-correlate the $2-5$ keV light curve with the $12-60$ keV 
light curve for different time bin sizes ranging from 5 ms to 20 s. For the bintime of 25 ms, we obtain 
a statistically significant distribution (i.e., high signal-to-noise ratio with small error-bars). We fit the peak (if any) 
with a Gaussian function to determine the peak position, and hence the time delay. 

\begin{figure*}
\includegraphics[scale=0.38,angle=-90]{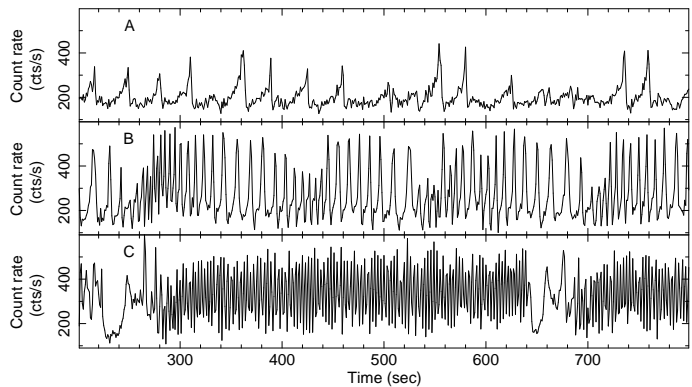}
\caption{{\it RXTE} PCA light curves ($2-60$ keV; 1 s binning) of 
IGR J17091--3624. Panel {\it A}: June 25, 2011 data ($\rho$ class); panel {\it B}: 
June 29, 2011 data (C1 class); and panel {\it C}: July 01, 2011 data (see \S~\ref{Dataanalysis}).
\label{c1lc}}
\end{figure*}

\section{Results}\label{Results}

\begin{figure*}
  \begin{center}
\includegraphics[scale=0.38,angle=-90]{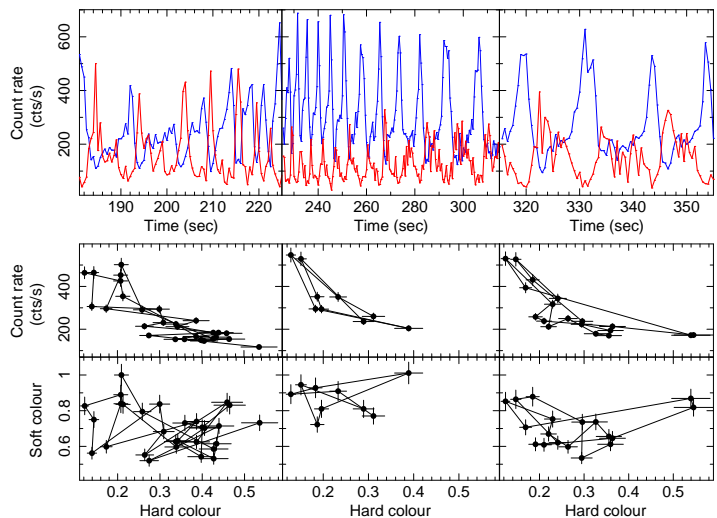}
\caption{{\it RXTE} PCA light curves, CD and HID of the C1 class of IGR J17091--3624.
{\it Top panels:} count rate (blue line) and hard colour (red line) with time.
The hard colour values are multiplied with 500 to bring them in the scale of count rates.
These adjacent panels show that the shape and frequency of the C1 class variability
evolve significantly in tens of seconds.
{\it Bottom panels:} HID and CD corresponding to each light curve segment of the 
top panels (see \S~\ref{Dataanalysis} and \ref{Results}).
\label{c1cd}}
\end{center}
\end{figure*}

\begin{figure*}
\begin{center}$
\begin{array}{ccc}
\includegraphics[scale=0.25,angle=-90]{fig3.eps} &
\includegraphics[scale=0.25,angle=-90]{fig4.eps} &
\includegraphics[scale=0.25,angle=-90]{fig5.eps} \\ 
\includegraphics[scale=0.22]{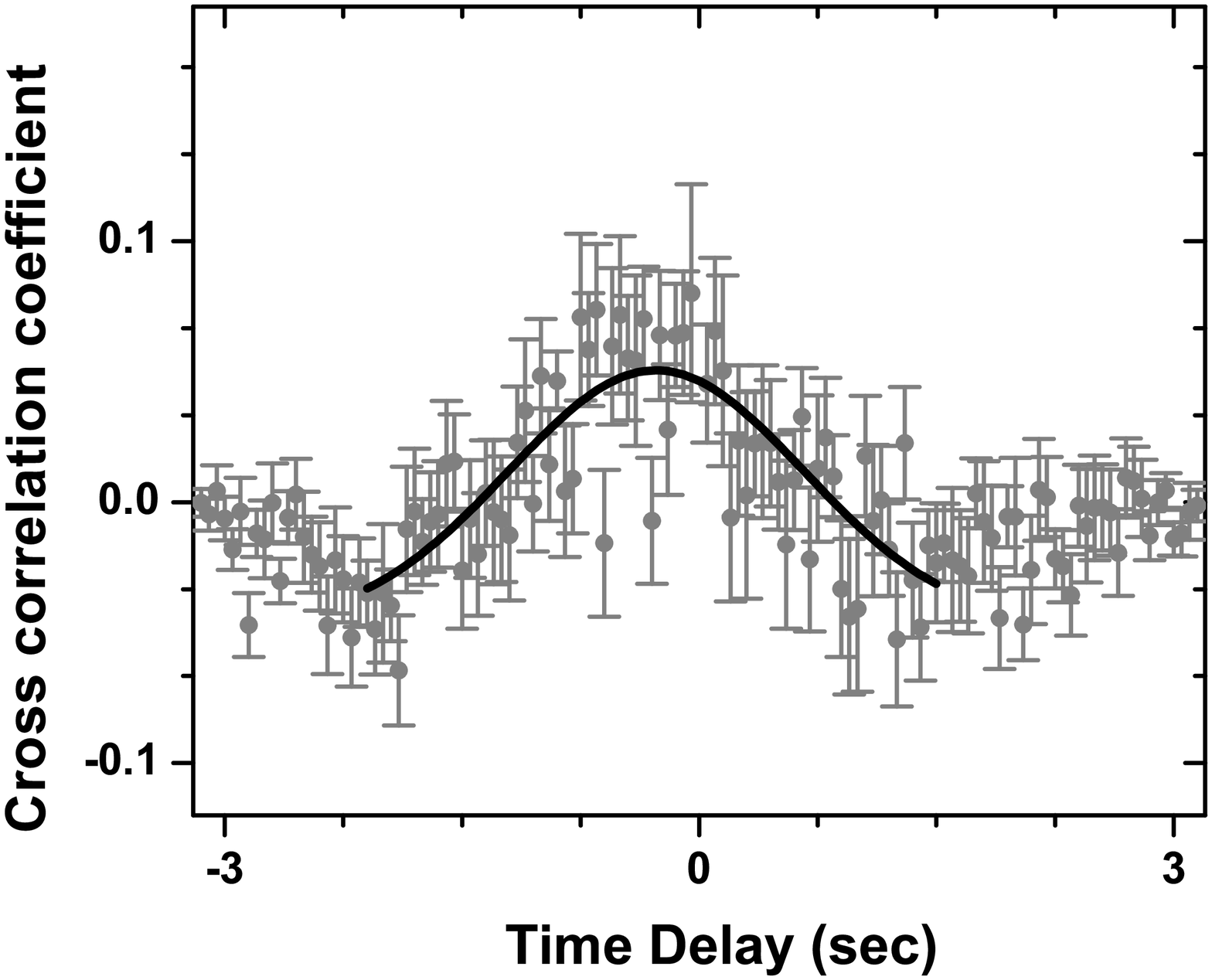} &
\includegraphics[scale=0.22]{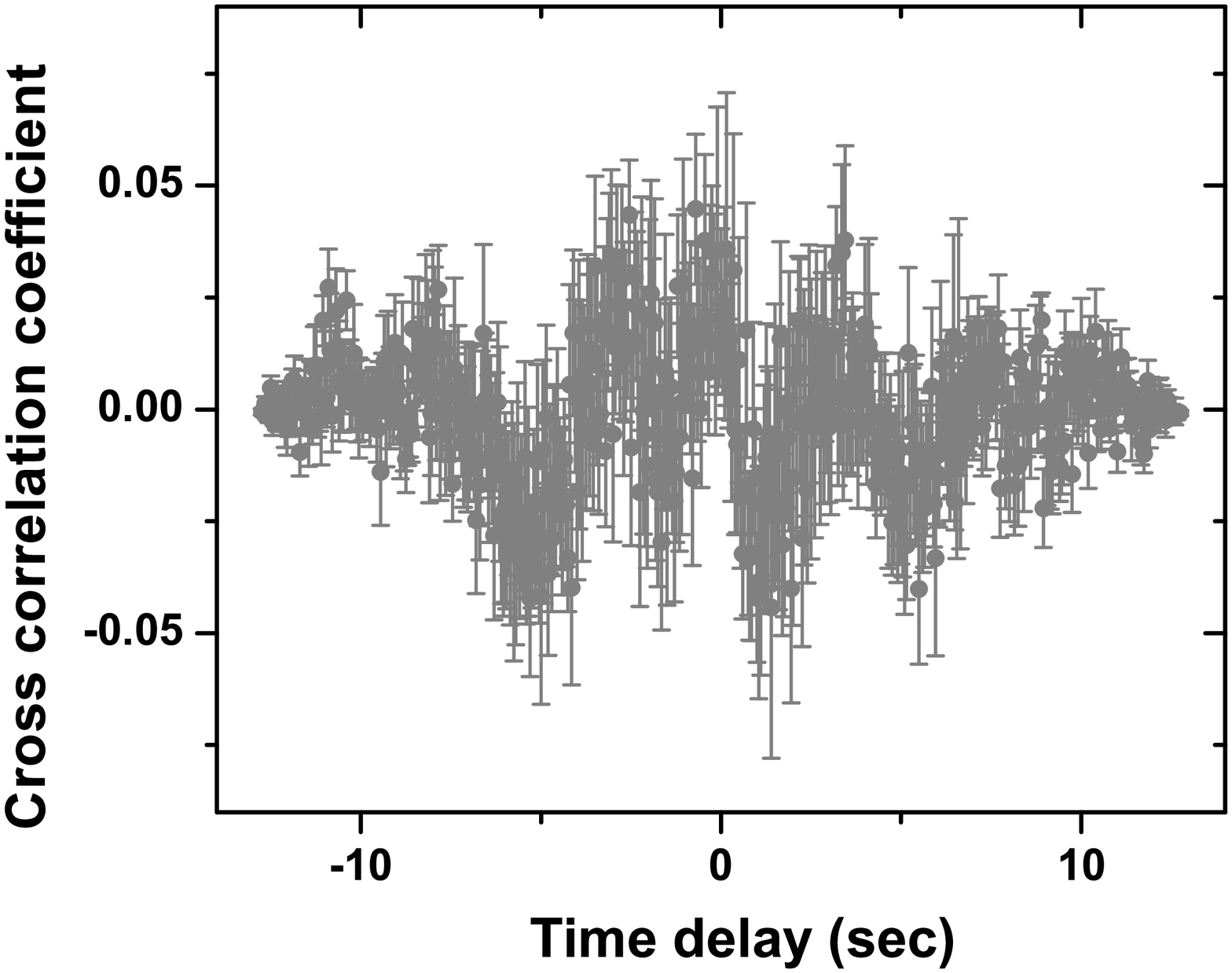} &
\includegraphics[scale=0.22]{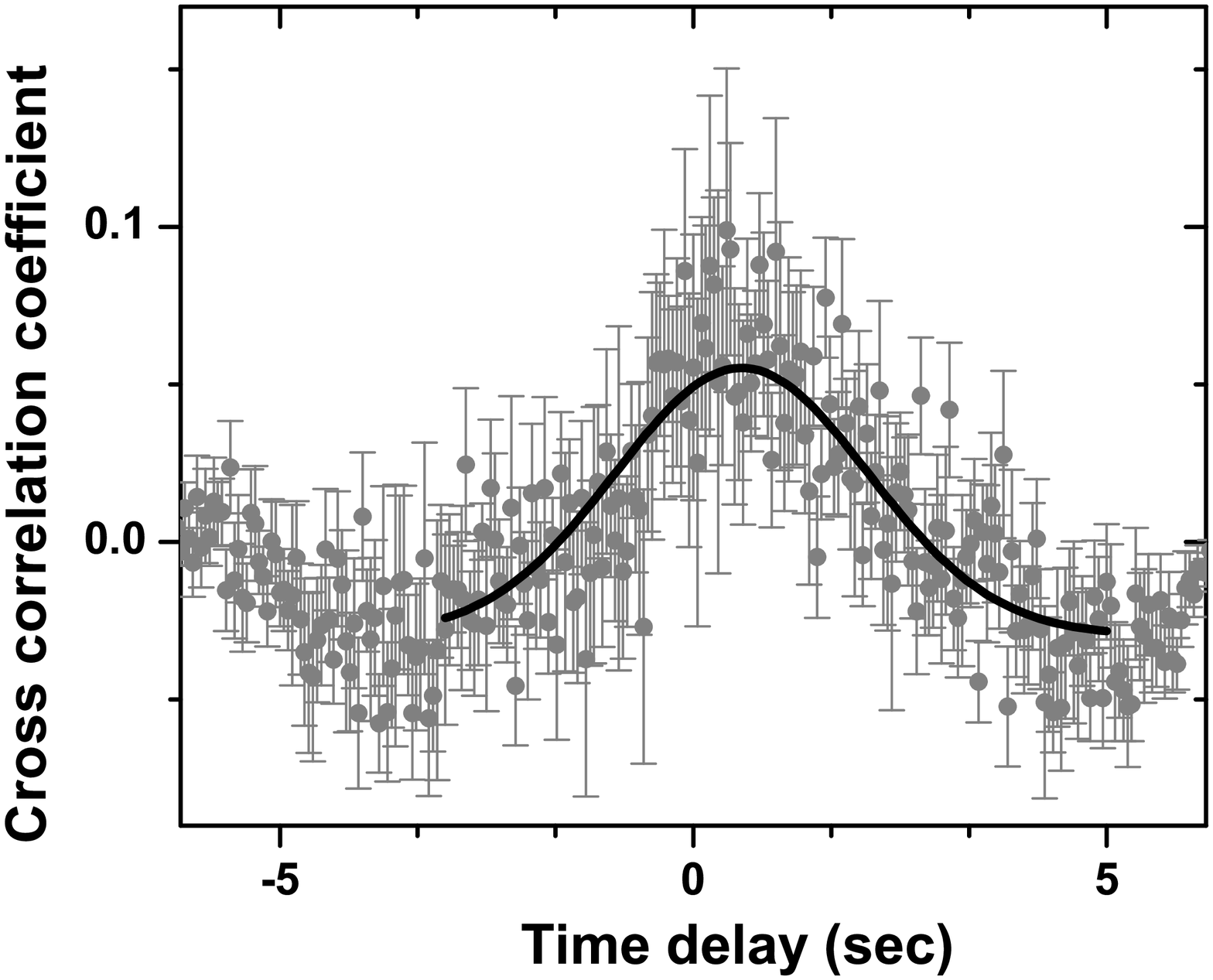} 
\end{array}$
\end{center}
\caption{{\it Top panels:} {\it RXTE} PCA rms normalized power density spectrum 
corresponding to each of the adjacent time segments (as shown in the top panels of 
Fig.~\ref{c1cd}) for IGR J17091--3624. The error bars are of 1$\sigma$ size, and the arrows 
indicate plausible QPOs.
{\it Bottom panels:} time delay measurements between the $2-5$ keV light curve and 
the $12-60$ keV light curve using cross correlation techniques for each of the adjacent 
segments shown in top panels of Fig.~\ref{c1cd}. The peaks for the first and the third segments
are fitted with a Gaussian function (see \S~\ref{Dataanalysis} and \ref{Results}).
\label{c1power}}
\end{figure*}

In this section, we discuss various properties (mentioned in \S~\ref{Dataanalysis}) of
the C1 class and the C2 class observed from IGR J17091--3624. 
The light curve of the C1 class is somewhat similar to that
of $\rho$ class \citep{Bellonietal2000,Paharietal2011}, but in this case, the $\rho$ class-like variabilities
are grouped in repetitive segments. There are other unique properties which
make the C1 class different from the $\rho$ class. For example, within one variability
group of the C1 class, even in less than 200 s, the shape and frequency of the 
intensity oscillations and the hard colour values evolve substantially. For example, in Fig.~\ref{c1cd}, 
during the first 45 s (first segment) the lightcurve has moderate oscillation frequency 
and relatively low peak intensity, while the hard colour values are relatively high during intensity dips.
During the next 80 s (second segment; Fig.~\ref{c1cd}), the oscillation frequency and intensity increase, 
while the hard colour values during intensity dips are relatively low. During the last 45 s
(third segment; Fig.~\ref{c1cd}),
the oscillation frequency decreases, the oscillation peak intensity remains relatively high,
and the hard colour values during intensity dips remain relatively low. The spectral evolution
in the relatively short time scale is also seen from the HIDs and CDs (Fig.~\ref{c1cd}), especially
from the hard colour range and the soft colour range populated for each segment. These types of spectral
and intensity evolutions in such a short time scale have never been observed for the $\rho$ class.
In the same short time scale, the power spectrum of C1 class also evolves, as indicated from
the appearance (in first and third segments) and the disappearance (in second segment) 
of a plausible quasi-periodic oscillation (QPO) near $2$ Hz with the significance of 2.9$\sigma$ and 2.6$\sigma$ respectively,
as well as the change of shape and total integrated power of the power spectrum in $0.05-10.0$ Hz
(Fig.~\ref{c1power}). Contrary to this, the power spectrum (including QPOs) of $\rho$ class
remains stable for much longer time. Finally, we find that the $2-5$ keV photons lag behind the
$12-60$ keV photons by $0.12\pm0.04$ s in the first 45 s of Fig.~\ref{c1cd} with the statistical accuracy of 2.9$\sigma$.
But in the last 45 s (third segment) of Fig.~\ref{c1cd}, the $12-60$ keV photons lag behind
the $2-5$ keV photons by $0.58\pm0.03$ s with the statistical accuracy of 18.5$\sigma$(see Fig.~\ref{c1power}). In case of $\rho$ class, this time delay
does not evolve. For the $\rho$ variabilities from GRS 1915+105, 
we find that the $12-60$ keV photons lag behind
the $2-5$ keV photons by a non-evolving value of $3.61\pm0.06$ s.
These show that the C1 class is different from the $\rho$ class, and is a unique class.
Besides, the repeatability of the C1 variability (June 28, 29 and July 20, 23, 24, 2011)
argues that it can be defined as a new class.
Fig.~\ref{c1lc} shows that the source evolved from the $\rho$ class into the 
C1 class. After that, it evolved into a $\beta$-type class with hard, quiescent dips but without soft dips (see panel {\it C} of Fig.~\ref{c1lc})
\citep{Altamiranoetal2011b,Bellonietal2000}.

\begin{figure*}
\includegraphics[scale=0.38,angle=-90]{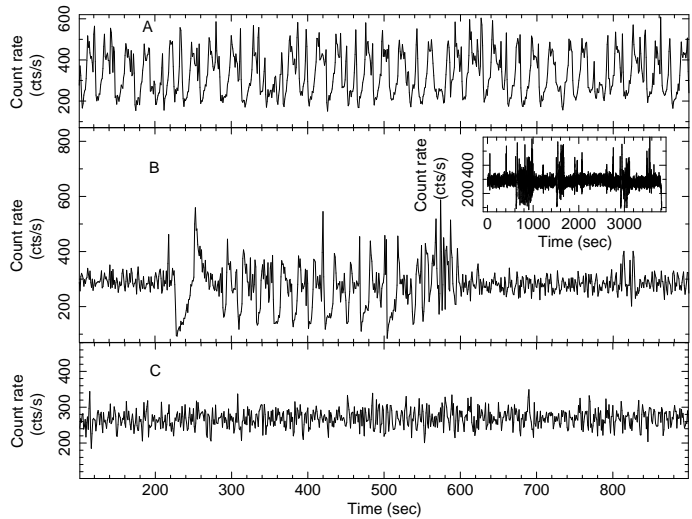}
\caption{{\it RXTE} PCA light curves ($2-60$ keV; 1 s binning) of
IGR J17091--3624. Panel {\it A}: September 20, 2011 data; panel {\it B}:
September 28, 2011 data (C2 class); and panel {\it C}: September 29, 2011 data. 
The inset of panel {\it B} shows the same C2 class
light curve for longer duration, so that several variable segments are seen
(see \S~\ref{Dataanalysis}).
\label{c2lc}}
\end{figure*}

\begin{figure*}
\includegraphics[scale=0.38,angle=-90]{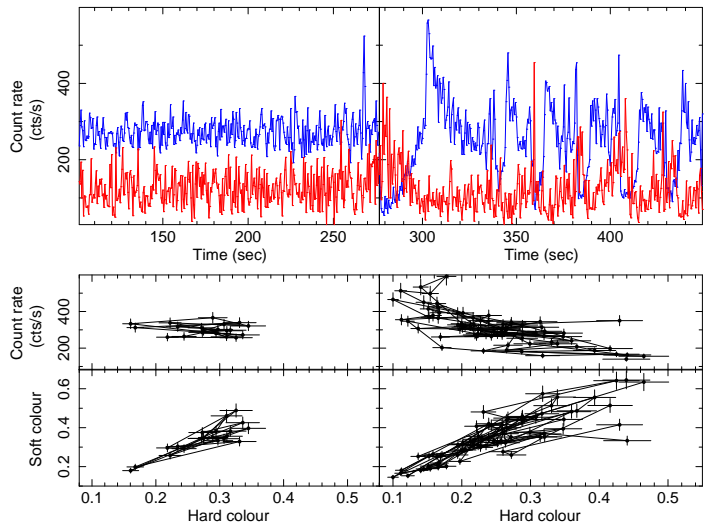}
\caption{{\it RXTE} PCA light curves, CD and HID of the C2 class of IGR J17091--3624.
{\it Top panels:} count rate (blue line) and hard colour (red line) with time.
The hard colour values are multiplied with 500 to bring them in the scale of count rates.
These adjacent panels show the natures of the non-variability and the variability
portions of the C2 class.
{\it Bottom panels:} HID and CD corresponding to each light curve segment of the
top panels (see \S~\ref{Dataanalysis} and \ref{Results}).
\label{c2cd}}
\end{figure*}

\begin{figure*}
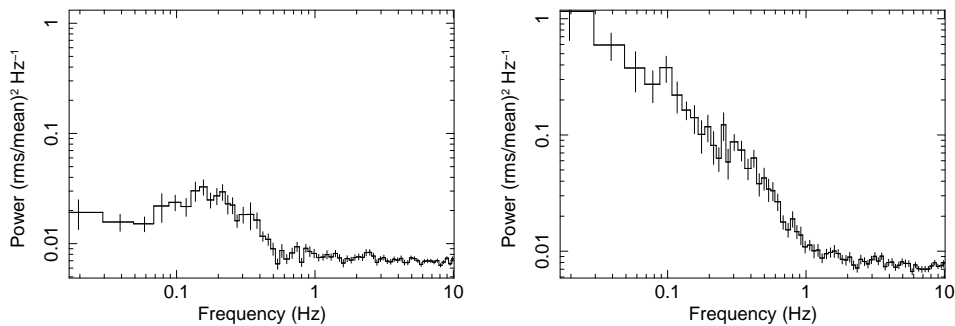

\begin{center}$
\begin{array}{cc}
\includegraphics[scale=0.25,angle=-90]{fig11.eps} &
\includegraphics[scale=0.25,angle=-90]{fig12.eps}
\end{array}$
\end{center}
\caption{{\it RXTE} PCA rms normalized power density spectrum
corresponding to each of the adjacent time segments (as shown in the top panels of
Fig.~\ref{c2cd}) for IGR J17091--3624. The error bars are of 1$\sigma$ size.
\label{c2power}}
\end{figure*}

The light curve of the C2 class alternately shows a nonvariable substate and a highly variable substate 
(Fig.~\ref{c2lc}). Light curves with such structures have never been seen from any 
black hole X-ray binary, not even from GRS 1915+105.
The highly variable substate can last for $\sim 300-400$ s, and the transition between
the two substates can happen as rapidly as in $\sim$ 5 s. The average intensities of the two substates
are similar to each other (Fig.~\ref{c2cd}). 
This figure shows that the transition from the nonvariable substate
to the highly variable substate happened without clear change of hard colour values, which 
is unusual and hardly seen in other black hole X-ray binaries. The mean values of hard colour
and soft colour are similar in both the substates (which is also unusual among black hole X-ray binaries), 
although these colours fluctuate more 
in the highly variable substate (see Fig.~\ref{c2cd}). Fig.~\ref{c2lc} shows that IGR J17091--3624
evolved from a strong (roughly uniform) variability dominated state into the C2 class variability,
and then it transitioned into a nonvariable state. This is unusual, because for GRS 1915+105,
a highly variable state transitions into a nonvariable state usually via a state having long quiescent dips.
The power spectra of the two subsets are different from each other (Fig.~\ref{c2power}).
The total integrated rms (in $0.05-10$ Hz) of the nonvariable state is at least 11 times lesser 
than that of the variable state, which is not very surprising. However, there is a
broad hump-like feature around 0.2 Hz in the PDS of the nonvariable state.
Based on the above unique properties of the C2 variability, as well as its repeatability
(August 25, 26 and September 21, 22, 23, 28, 2011), it can be defined as a new class.

\section{Discussion and Conclusions}\label{Discussion}

IGR J17091--3624 is the only black hole X-ray binary after GRS 1915+105,
which showed a few types of large intensity variabilities.
Some of these are similar to those seen from the GRS 1915+105. 
Why is the discovery of new variability classes important? As \citet{Bellonietal2000}
pointed out, this gives the chance to look for the basic `states' of the source,
and to have complete details about the source. Besides, the classes have been
modeled to probe the physics of accretion-ejection processes in case of
the accreting black hole GRS 1915+105 (e.g., \citet{Neilsenetal2011}).
Therefore, our finding of two new variability classes will be useful 
to obtain detailed understanding of IGR J17091--3624 and GRS 1915+105,
as well as to probe the inflow-outflow mechanisms of accreting black holes.

The main purpose of this Letter is to report the discovery of two new variability
classes C1 and C2. An attempt to probe their physical origins and to discuss
their implications will be taken up in future detailed studies. Nevertheless,
here we list, what could be learn from these new classes.
(1) In case of GRS 1915+105, it is observed that most of the highly variable states are 
stabilized (i.e., become non-variable) via long, nonvariable and quiescent dips 
(like hard dips in $\beta$ class, long quiescence in $\alpha$ class, etc.). 
However, C2 class is an exception to this, where strong variabilities transition into 
a nonvariable state with similar average intensities, and long quiescent dips, before and after 
strong variabilities are not observed. This finding will have impact on models, which attempt
to explain class transitions.
(2) For GRS 1915+105, average hard colour decreases when the source transitions from 
a nonvariability into a variability state \citep{Yadavetal1999}. Besides, the variability from this source is 
always found in soft state. Contrary to this,
In case of C2 class of IGR J17091--3624, both variable and 
nonvariable substates have similar mean hard colour,
and transition from one substate to another does not require clear hard colour change.
These show that variability does not necessarily require a different spectral state (for example, soft
thermal state) from the spectral states of nonvariability, and hence this will have impact
on the variability models.
(3) The finding of rapid evolution of spectral and timing properties during C1 class
will also be useful to constrain the variability models \citep{Mineoetal2012, Neilsenetal2011}. Especially the change of soft lag into
hard lag in tens of seconds could be useful to constrain the properties (e.g., location, size)
of spectral components. Such a time delay reversal may indicate a strong coupling among
the spectral components.
(4) Finally, the disappearance and reappearance of a plausible $\sim 2$ Hz QPO in tens of seconds 
for C1 class may have impact on the models of accretion disc instabilities. Note that
these instabilities are widely believed to be responsible for the variabilities of
GRS 1915+105 \citep{Bellonietal2000}.

\section{Acknowledgements}

This research has made use of data obtained through the High Energy Astrophysics Science Archive Research Center online service, provided by the NASA/Goddard Space Flight Center.

\bsp
\label{lastpage}

\end{document}